# Probing the Microscopic Origin of Gravity via Precision Polarization and Spin Experiments

Wei-Tou Ni

Purple Mountain Observatory

No. 2, Beijing W. Rd., Nanjing, 210008 China

e-mail: wtni@pmo.ac.cn

## ABSTRACT

As in other parts of physics, we advocate the interaction approach: experiments ←→ phenomenology ←→ low-energy effective (field) theory ←→ microscopic theory to probe the microscopic origin of gravity. Using χ-g phenomenological framework, we discuss the tests of equivalence principles. The only experimentally unconstrained degree of freedom is the axion freedom. It has effects on the long-range astrophysical/cosmological propagation of electromagnetic waves and can be tested/measured using future generation of polarization measurement of cosmic background radiation. The verification or refutal of this axionic effect will be a crucial step for constructing effective theory and probing the microscopic origin of gravity. The interaction of spin with gravity is another important clue for probing microscopic origin of gravity. The interplay of experiments, phenomenology and effective theory is expounded. An ideal way to reveal the microscopic origin of gravity is to measure the gyrogravitational ratio of particles. Three potential experimental methods are considered.

## I. INTRODUCTION

The discovery that the expansion of our Universe is accelerating and the cosmological constant is nonzero put general relativity into an empirically tight position. Semi-classical or microscopic gravitation theory is needed --- the cosmological constant needs to be explained and the microscopic origin of gravity needs to be probed. As in other parts of physics, we advocate the interaction approach: experiments ←→ phenomenology ←→ low-energy effective (field) theory ←→ microscopic theory. Although experimentation is difficult, there are a few directions that precision experiments can be performed to reveal the microscopic origin of gravity. We first use $\chi$-g phenomenological framework to discuss the tests of equivalence principles and the axion freedom. The axion freedom has effects on the long-range astrophysical/cosmological propagation of electromagnetic waves and can be tested/measured using future generation of polarization measurement of cosmic background radiation. The interaction of spin with gravity is another important aspect to probe the microscopic origin of gravity. For this we consider Naik-Pradhan theory and Arkani-Hamed-Cheng-Luty-Mukohyama effective field theory with ghost condensation, and compare them with experiments. The meaning of measuring the gyrogravitational ratio of particles is explained and potential ways to measure it are considered.

II. $\chi$-g FRAMEWORK, AXIAL INTERACTION AND POLARIZATION

The $\chi$-g framework for the gravitational coupling of the electromagnetism [1-3] can be summarized in the following interaction Lagrangian density:

$$L_I = -(1/(16\pi))\chi^{ijkl} F_{ij} F_{kl} - A_k j^k (-g)^{(1/2)} - \Sigma_I m_I (ds_I)/(dt) \delta(\mathbf{x}-\mathbf{x}_I), \tag{1}$$

where $\chi^{ijkl} = \chi^{klij} = -\chi^{klji}$ is a tensor density of the gravitational fields (e.g., $g_{ij}$, $\varphi$, etc.) or fields to be investigated and $j^k$, $F_{ij} \equiv A_{j,i} - A_{i,j}$ have the usual meaning. The gravitation constitutive tensor density $\chi^{ijkl}$ dictates the behavior of electromagnetism in a gravitational field and has 21 independent components in general. For a metric theory (when EEP holds), $\chi^{ijkl}$ is determined completely by the metric $g^{ij}$ and equals $(-g)^{1/2} [(1/2) g^{ik} g^{jl} - (1/2) g^{il} g^{kj}]$. In the phenomenological approach, the structure of $\chi^{ijkl}$ is to be constrained by experiments and observations. The null-birefringence observations of pulses and micropulses of pulsars give strict constraints on $\chi^{ijkl}$. The condition for no splitting (no birefringence, no retardation) of pulses in all directions gives ten constraints on the $\chi^{ijkl}$. With these ten constraints, $\chi^{ijkl}$ can be written in the following form

$$\chi^{ijkl} = (-H)^{1/2} [(1/2) H^{ik} H^{jl} - (1/2) H^{il} H^{kj}] \psi + \varphi e^{ijkl}, \tag{2}$$

where $H_{ij}$ is a metric and $H = \det(H_{ij})$ [2, 3]. From pulsar data, these relations are empirically verified to better than $10^{-15}$ in our galaxy. Data from measurements of extragalactic radio sources constrain these relations to much better accuracy in extragalactic and cosmological distances [4]. The electromagnetic propagation in Moffat's nonsymmetric gravitational theory fits the $\chi$-g framework [5] and can be constrained stringently. The effect of $\varphi$ in (2) is to change the phase of two different circular polarizations of electromagnetic-wave propagation in gravitation field and gives polarization rotation for linearly polarized light [6-8]. Polarization observations of radio galaxies put a limit of $\Delta\varphi \leq 1$ over cosmological distance [9-14]. Further observations to test and measure $\Delta\varphi$ to $10^{-6}$ is promising. The natural coupling strength $\varphi$ is of order 1. However, the isotropy of our observable universe to $10^{-5}$ may leads to a change $\Delta\varphi$ of $\varphi$ over cosmological distance scale $10^{-5}$ smaller. Hence, observations to test and measure $\Delta\varphi$ to $10^{-6}$ are needed. In 2002, DASY microwave interferometer observed the polarization of the cosmic background. With the axial interaction (2), the polarization anisotropy is shifted relative to the temperature anisotropy. In 2003, WMAP found that the polarization and temperature are correlated. This gives a constraint of $10^{-1}$ of $\Delta\varphi$. Planck Surveyor will be launched in 2007 with better polarization-temperature measurement and will give a sensitivity to $\Delta\varphi$ of $10^{-2}$-$10^{-3}$. A dedicated future experiment on cosmic microwave background radiation will reach $10^{-5}$-$10^{-6}$ $\Delta\varphi$-sensitivity. This is very significant as a positive result may indicate that our patch of inflationary universe has a 'spontaneous polarization' in fundamental law of electromagnetic propagation influenced by neighboring patches (Fig. 1) and we can 'observe' neighboring patches through a determination of this fundamental physical law; if a negative result turns out at this level, it may give a good constraint on superstring theories as axions are natural to superstring theories.

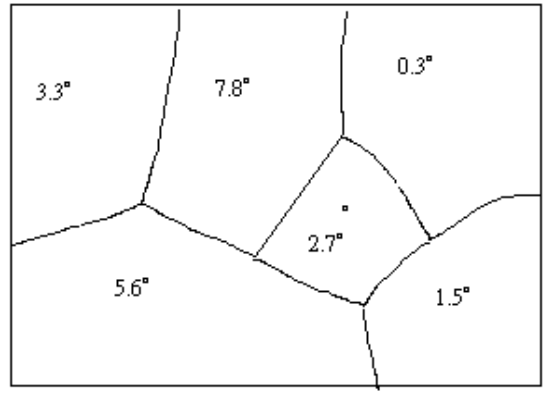

Fig. 1. Inflationary universes. In this picture, our universe occupies a small region in the 2.7 K patch. We may be influenced by other inflationary patches to have a 'spontaneous polarization' in fundamental law of electromagnetic propagation and we can 'observe' neighboring patches through the determination of this fundamental physical law.

As to $H_{ij}$ and $\psi$ in Eq.(2), Hughes-Drever experiments and free-fall experiments fix them to the metric form with a high accuracy. This restricts (2) to

$$\chi^{ijkl}=(-g)^{1/2}\,[(1/2)\,g^{ik}\,g^{jl} - (1/2)\,g^{il}\,g^{kj}] + \varphi\,e^{ijkl}. \qquad (3)$$

with a high accuracy. (3) is the axion theory for electromagnetism [1, 6]. The phenomenological Lagrangian (1) links with effective field theory Lagrangians fittingly [15] and passes empirical constraints to them.

III. THE ROLE OF SPIN IN GRAVITATION

Spin (helicity in the case of zero mass) and mass are 2 independent Casimir invariants of the Poincare group. What is the role of spin in gravitation? Mach's principle and effective field theory motivated the investigations of cosmic spatial isotropy for particles with spin. Recent experiment shows that the spin-dependent cosmic anisotropy for electron is less than $3.1 \times 10^{-20}$ eV [16]. Not surprisingly, this experiment also gives stringent constraints on the Bluhm-Kostelecky effective field theory of Lorentz and CPT violations: the 2 violating parameters $b^e_\perp$ and $b^e_Z$ are constrained to be less than $3.1 \times 10^{-29}$ GeV and $7.1 \times 10^{-28}$ GeV respectively.

Gauging a subgroup of Lorentz group, Naik and Pradhan [17] introduce a massless axial vector gauge field, axial photon which gives rise to a super-weak long-range spin-spin force between two particles:

$$V(\underline{r}) = - (g^2/2r)\{ (\underline{S}_1 \cdot \underline{S}_2) + [(\underline{S}_1 \cdot \underline{r})(\underline{S}_2 \cdot \underline{r})/r^2]\}, \qquad (4)$$

where $g$ is a coupling constant and $\underline{S}_1$, $\underline{S}_2$ are the spins of two particles. From a precise experiment [18], the strength of anomalous spin-spin interaction is constrained to $(1.2 \pm 2.0) \times 10^{-14}$ of the magnetic spin-spin interaction, and this limits the coupling of axial photon to a level much lower than originally proposed.

From dimensional argument, a spin-spin interaction can have the following form:

$$V(\underline{r}) = (g^2/2r)\{K_1 (\underline{S}_1 \cdot \underline{S}_2) + K_2 [(\underline{S}_1 \cdot \underline{r})(\underline{S}_2 \cdot \underline{r})/r^2]\}, \tag{5}$$

where $K_1$ and $K_2$ are constants. In cosmology, there is a preferred frame. However, in general relativity, there is no specific preferred frame. From this consideration, modern cosmological observations and field theory development, Arkani-Hamed, Cheng, Luty and Mukohyama [19] proposed an effective field theory of gravity with ghost condensation which gives infrared modification of gravity and has impirically testable predictions. If the standard model fields have direct coupling with the ghost sector, there are spin couplings. Between particle with spins, there is an interaction of the form

$$V(\underline{r}) \sim (M^4/M'^2F^2)(1/r)\{ (\underline{S}_1 \cdot \underline{S}_2) - 3 [(\underline{S}_1 \cdot \underline{r})(\underline{S}_2 \cdot \underline{r})/r^2]\}, \tag{6}$$

where $M^4/M'^2F^2$ is the coupling constant. (6) has the form of (5) with $K_1 = 1$ and $K_2 = -3$. This gives an effective field theory link for the phenomenological Lagrangian (5). From the experiment [18], the coupling constant $M^4/M'^2F^2$ is constrained to be less than $\sim 10^{-42}$. The effective field theory [19] can induce long-range oscillatory behavior like MOND (Modified Newtonian Dynamics) [20], and therefore can be tested by precision solar-dynamics missions like ASTROD I [21]. The 'invisible' axion theory can also be tested and constrained by experiments [22].

Ten years after the discovery of general relativity, in 1925-26, Goudsmit and Uhlenbeck [23] introduced our present concept of electron spin. Standard way to incorporate spin into the classical general relativity is to treat the aggregate of spins as ordinary angular momentum. However, as we know, for the electromagnetic interaction, the gyromagnetic ratios of elementary particles are different from one, and these ratios reveal the inner electromagnetic structures of elementary particles. What would be the gyrogravitatinal ratios of elementary particles? If they differ from one, they will definitely reveal the inner gravitational structures of elementary particles. These will give clues to the microscopic origin of gravity.

Although experimental measurement of gyrogravitational ratios of particles are difficult, the following 3 methods have potentials:

(i) Polarized-body method [24],

(ii) Atom interferometry [25],

(iii) Superfluid $^3$He [26].

IV. OUTLOOK

The interaction approach: experiments ←→ phenomenology ←→ low-energy effective (field) theory ←→ microscopic theory for probing the microscopic approach is illustrated. It has been useful and will be fruitful in this endeavor, as in the history of the development of gravitation [27].